\documentclass[aps,prb,floatfix,amsmath,amssymb,eqsecnum,nofootinbib,twocolumn,superscriptaddress]{revtex4}

\usepackage{amsmath,amssymb,dsfont,graphicx,psfrag}
\usepackage{epsfig}
\usepackage{graphicx}

\def\beq{\begin{equation}}
\def\eeq{\end{equation}}
\def\beqr{\begin{eqnarray}}
\def\eeqr{\end{eqnarray}}

\begin{document}
\title{Emergence of helical edge conduction in graphene at the $\nu=0$ quantum Hall state}

\author{Pavel Tikhonov}
\affiliation{Department of Physics, Bar-Ilan University, Ramat-Gan 52900, Israel}

\author{Efrat Shimshoni}
\affiliation{Department of Physics, Bar-Ilan University, Ramat-Gan 52900, Israel}

\author{H.~A.~Fertig}
\affiliation{Department of Physics, Indiana University, Bloomington, IN 47405, USA}

\author{Ganpathy Murthy}
\affiliation{Department of Physics and Astronomy, University of Kentucky, Lexington KY 40506-0055, USA}

\date{\today}
\begin{abstract}
The conductance of graphene subject to a strong, tilted magnetic field exhibits a
dramatic change from insulating to conducting behavior with tilt-angle, regarded as evidence for the
transition from a canted antiferromagnetic (CAF) to a ferromagnetic
(FM) $\nu=0$ quantum Hall state. We develop a theory for the electric transport in this system based on the spin-charge connection, whereby the evolution in the nature of collective spin excitations is reflected in the charge-carrying modes. To this end, we derive an effective field theoretical description of the low-energy excitations, associated with quantum fluctuations of the spin-valley domain wall ground-state configuration which characterizes the two-dimensional (2D) system with an edge. This analysis yields a model describing a one-dimensional charged edge mode coupled to charge-neutral spin-wave excitations in the 2D bulk. Focusing particularly on the FM phase, naively expected to exhibit perfect conductance, we study a mechanism whereby the coupling to these bulk excitations assists in generating back-scattering. Our theory yields the conductance as a function of temperature and the Zeeman energy - the parameter that tunes the transition between the FM and CAF phases - with behavior in qualitative agreement with experiment.

\end{abstract}
\maketitle

\section{Introduction and Principal Results}
\label{sec:intro}

One of the most intriguing manifestations of many-body effects in graphene is the observation of a quantum Hall (QH) state at $\nu=0$ in the presence of strong perpendicular magnetic fields \cite{Zhang_Kim2006,Alicea2006,Goerbig2006,Gusynin2006,Nomura2006,Jiang2007,Herbut2007,
  Fuchs2007,Abanin2007,Checkelsky,Du2009,Goerbig2011,Dean2012,Yu2013}.
This unique state is characterized by a plateau at $\sigma_{xy}=0$, and a peak in the longitudinal resistance which typically exhibits insulating behavior. The high resistance signature is difficult to reconcile with a non-interacting theory \cite{Abanin_2006}, which implies a helical nature of the edge states: right and left movers have opposite spin flavors, resolved by the Zeeman splitting of the $n=0$ Landau level in the bulk. In analogy with the
quantum spin Hall (QSH) state in two-dimensional (2D) topological
insulators \cite{Kane-Mele,TIreview}, the edge states are hence immune to backscattering by static impurities, and a nearly perfect conduction is expected.

Coulomb interactions do not change the character of the edge states in a fundamental way, as long as the many-body state forming in the bulk remains spin-polarized, i.e. is a ferromagnet (FM). Such a bulk phase supports a gapless collective edge mode associated with a domain wall in the spin configuration, which can be modeled as a helical Luttinger liquid \cite{Fertig2006,SFP,Paramekanti,Kusum}. Insulating behavior therefore suggests that the true ground state is not a FM. Indeed,
at half filling of the $n=0$ Landau level, there is a rich variety of ways to spontaneously break the $SU(4)$ symmetry in spin and
valley space, leading to a multitude of possible ground states with distinct properties
\cite{Herbut2007AF,Jung2009,Nandkishore,Kharitonov_bulk,Kharitonov_edge,SO5,Roy2014,Lado2014,QHFMGexp}.
The combined effect of interactions and external fields can assist in selecting the favored many-body ground state, particularly when accounting for lattice-scale interactions which do not obey $SU(4)$ symmetry.
Most interestingly, the tuning of an external parameter can drive a transition from one phase to another.
As a concrete example, it has been proposed \cite{Kharitonov_bulk,Kharitonov_edge,bilayer_QHE_CAF} that a phase transition can occur from a canted
antiferromagnetic (CAF) to a FM state, tuned by increasing the Zeeman energy $E_z$ to appreciable values.

Recent experiments in a tilted magnetic field
\cite{Young2013,Maher2013} appear to confirm the predicted phase
transition in a transport measurement. In these experiments, the
perpendicular field $B_{\perp}$ is kept fixed while the Zeeman
coupling $E_z$ is tuned by changing the parallel component. At
$\nu=0$ and relatively low $E_z$, the system exhibits a vanishing
two-terminal conductance which slightly increases to finite values
with increasing temperature $T$; i.e., it indicates an insulating
behavior as in earlier studies of the $\nu=0$ state. However with
increasing values of $E_z$, the sample develops a steep rise of
conductance and approaches an almost perfect two-terminal
conductance of $G\approx 2e^2/h$, a behavior characteristic of a
QSH state with protected edge states.

The most natural interpretation of these findings is in terms of
the predicted phase transition from a CAF to a FM bulk state.
However, while the theory dictates a second order quantum phase
transition at a critical Zeeman coupling $E_z^c$ (and $T=0$),
the transport data (obtained at finite $T$) reflects a smooth
evolution of $G$ with $E_z$. The critical point $E_z^c$ can be
estimated only roughly by, e.g., identifying the value of $E_z$ where
$G(T\rightarrow 0)$ approaches the mid-value $e^2/h$, or where
$dG/dT$ changes sign. At the highest accessible $E_z$ (where
presumably $E_z>E_z^c$), the conductance still falls below the
perfect quantized value.

The above described behavior suggests that the low energy
charge-carrying excitations smoothly evolve through the CAF-FM
phase transition, so that their change of character reflects the
critical properties of the bulk phases. In earlier work \cite{MSF2014,tdhfa}, we
showed that in both phases one can construct collective charged modes associated with
textures in the spin and valley configurations near the edges of the system, and characterized their essential properties.
Such excitations are supported due to the formation of a domain wall (DW) structure, where the
spin and valley are entangled and vary with position towards the edge. The nature of collective edge modes continuously evolves as $E_z$ is tuned through the transition. In particular, the CAF phase
supports a gapped charged edge mode, which becomes gapless at the transition to the FM phase and is smoothly connected to the helical edge mode characteristic of the QSH state.

In terms of the spin degree of freedom, the gapless charged collective edge mode in the FM phase corresponds to a
$2\pi$ twist of the ground-state spin configuration in the
$XY$-plane\cite{Fertig2006}. This spin twist is imposed upon the
spatially-varying $S_z$ associated with the DW, thus creating a spin
texture (i.e. a Skyrmion stretched out along the entire edge), with an associated charge that is inherent to quantum Hall
ferromagnets \cite{QHFM,Fertig1994,Yang2006}.
In contrast, the energy cost of generating such a spin texture in the CAF phase is infinite. A proper description of the lowest energy charged excitations in this phase therefore involves a coupling between topological structures at
the edge and in the bulk
\cite{MSF2014}, and yields a charge gap on the edge that encodes the {\it bulk} spin stiffness for rotations in the
$XY$-plane.

In both the CAF and FM phases, the collective excitations also contain charge-neutral modes, and among them the low-energy ones are spin-waves in the bulk \cite{tdhfa}. Their behavior across the transition is the opposite of the charged edge modes: in the CAF phase, where the charged edge excitations are gapped, a broken $U(1)$ symmetry in the bulk (associated with $XY$-like order
parameter) implies a neutral, gapless Goldstone mode. In contrast, in the FM phase where the charged edge mode is gapless, the bulk spin-waves acquire a gap which grows with $(E_z-E_z^c)$. While the neutral modes do not contribute to electric transport as carriers, their coupling to the charged modes can play an important role in the scattering processes responsible for a finite resistance. Most prominently, in the FM phase where the helical edge modes are protected by conservation of the spin component $S_z$, the coupling to the bulk spin-waves is essential to relax this conservation, and therefore dominates the electric resistance at finite $T$.

In a previous work\cite{tdhfa}, three of the present authors carried
out a detailed time-dependent Hartree-Fock (TDHF) analysis of the HF
state of our first paper\cite{MSF2014}. TDHF is similar in spirit to a
spin-wave analysis, in that it diagonalizes the Hamiltonian in the
Hilbert space of a single particle-hole excitation. However, for our
present purpose of investigating the transport on the edge near the
transition, we need to go beyond TDHF in several ways. Firstly, we
need to include a coupling between the edge and bulk modes that allows
the relaxation of the edge spin, which is otherwise a good quantum
number. Secondly, we need to introduce disorder at the edge, which is
extremely hard to do in TDHF. Thirdly, we would like the
temperature-dependence of transport coefficients close to the
transition to compare to experiments.

To accomplish these objectives, in this paper we will first derive a
low-energy effective field-theoretic description of the coupled system
of bulk and edge, which encodes the information on the nature of the
collective modes as well as the symmetries of the problem (overall
$S_z$ conservation, including both bulk and edge). The parameters
appearing in this effective theory have to be matched with the results
of TDHF as well as physical constraints such as the fact that the
stiffness is not singular at the transition. Since we focus on the
low-energy sector, the theory contains the charge-carrying edge mode
(gapless in the FM phase) and neutral
spin-wave excitations of the bulk (gapped
in the FM phase). Interestingly, some of the parameters of the
effective theory do behave in a singular way as the transition in
approached, reflecting a divergent length scale.

This effective theory contains all the ingredients we need to compute
transport coefficients
at low
temperatures. The detailed TDHF calculation\cite{tdhfa} shows that all
other collective excitations are high in energy, and remain gapped
through the transition. They will thus contribute, at best, to a
finite renormalization of the parameters of the effective theory.


Focusing particularly on the FM phase, we study the mechanism whereby the coupling of the charged edge mode to the charge-neutral bulk excitations assists in generating back-scattering. Our theory yields the two-terminal conductance $G$ as a function of $T$ and the Zeeman energy $E_z$. The main results are summarized in Fig. \ref{fig:Conductance}, and Eq.  (\ref{eq:deltaR2T_short}) below which describes the intrinsic resistance (dictating the deviation of $G$ from $2e^2/h$) as a scaling function of $T$ and the critical energy scale $\Delta=E_z-E_z^c$. In the low $T$ limit where $T\ll\Delta$, this yields a simple activation form [see Eq. (\ref{eq:deltaR_final})]. This behavior is dual to the exponentially small {\it conductance} expected in the insulating CAF phase.  Our results are in qualitative agreement with experiment.

The paper is organized as follows. In Sec. \ref{sec:2Daction} we detail the derivation of a 2D field theoretical model for the quantum fluctuations in the spin and valley configuration for a system with an edge potential. In Sec. \ref{sec:normalmodes} we study the normal modes of low energy collective excitations in the FM phase, and derive an effective Hamiltonian describing the 1D edge mode coupled to 2D bulk spin-waves. This section is supplemented by Appendix  \ref{sec:uK_critical}, devoted to a derivation of the scaling of the model parameter when $E_z$ approaches the critical value $E_z^c$. Based on the resulting effective model, in Sec. \ref{sec:G} we evaluate the two-terminal conductance $G$ as a function of $T$ and $E_z$.  Some further details of the calculation are included in Appendix \ref{sec:deltaRdetails}. Finally, our main results and some outlook are summarized in Sec. \ref{sec:summary}.

\section{Model for Spin-valley fluctuations in two-dimensions}
\label{sec:2Daction}

We consider a ribbon of monolayer graphene in the $x-y$ plane,
subject to a tilted magnetic field of magnitude $B_T$ and
perpendicular component $B_\perp$. These two distinct field scales
independently determine the Zeeman energy $E_z\propto B_T$ and the
magnetic length $\ell=\sqrt{\hbar c/e B_\perp}$. At zero doping,
the $n=0$ Landau level is half-filled and we assume that mixing
with other Landau levels can be neglected. In addition, for the
time being we focus on an ideal system uniform in the
$\hat{y}$-direction but of finite width in the
$\hat{x}$-direction, so that single-electron states can be labeled
by a guiding-center coordinate $X=\ell^2k_y$ with $k_y$ the
momentum in the $y$-direction. Similarly to Ref.
\onlinecite{MSF2014}, the boundaries of the ribbon are accounted
for by an edge potential $U(x)\hat{\tau}_x$ where $\hat{\tau}_x$
denotes a valley isospin operator, and $U(x)$ grows linearly over
a length scale $w$, from zero in the bulk to a constant $U_e$ on
the edge. It is therefore convenient to represent electronic
states in a basis of 4-spinors $|X s\,\tau \rangle$ where
$s=\uparrow,\downarrow$ denotes the real spin index $s_z$, and
$\tau=\pm$ are the eigenvalues of $\hat{\tau}_x$ corresponding to
symmetric and antisymmetric combinations of valley states.

The microscopic Hamiltonian describing the system, projected into the above manifold of $n=0$ states, assumes the form \cite{Kharitonov_bulk,Kharitonov_edge,MSF2014}
\begin{widetext}
\begin{eqnarray}\label{Hmicro}
H &=& \sum_X c^\dagger(X)
[-E_z\sigma_z\tau_0+U(X)\sigma_0\tau_x] c(X) +H_{int}\; ,\\
H_{int}&=&\frac{\pi\ell^2}{L^2}\sum_{\alpha=0,x,y,z}\sum_{X_1,X_2,q}
e^{-q^2\ell^2/2+iq(X_1-X_2)}
g_\alpha :c^\dagger(X_1+{{q\ell^2} \over 2})\tau_\alpha c(X_1-{{q\ell^2} \over 2})
c^\dagger(X_2-{{q\ell^2} \over 2})\tau_\alpha c(X_2+{{q\ell^2} \over 2}):, \nonumber
\end{eqnarray}
\end{widetext}
where $c^\dagger(X),c(X)$ are creation and annihilation operators written as 4-spinors,
[$c^\dagger(X) \equiv (c_{K,\uparrow}^{\dag}(X),c_{K,\downarrow}^{\dag}(X),
c_{K',\uparrow}^{\dag}(X),c_{K',\downarrow}^{\dag}(X))$],
$\sigma_\alpha$ ($\tau_\alpha$) are the spin (isospin) Pauli matrices and
$\sigma_0$, $\tau_0$ are unit matrices, $L$ is the system size and
$:\,:$ denotes normal ordering; $g_\alpha$ denote lattice-scale interaction parameters obeying $g_x=g_y \equiv g_{xy}$ and $g_z>-g_{xy}>0$. The latter condition is required \cite{Kharitonov_bulk} to stabilize a CAF phase for small $E_z$. Finally, $g_0$ parametrizes an $SU(4)$ symmetric interaction which mimics the effect of Coulomb interactions, and dominates the spin-isospin stiffness.

As we have shown in Ref. \onlinecite{MSF2014}, for arbitrary $E_z$ and $U(X)$ the Hartree-Fock solution of the Hamiltonian Eq. (\ref{Hmicro}) at $1/2$-filling is a spin-valley entangled domain wall, characterized by two distinct canting angles $\psi_a(X)$, $\psi_b(X)$ which vary continuously as a function of $X$ when approaching an edge. This corresponds to a Slater determinant with
two (out of four possible) occupied states for each $X$:
\begin{eqnarray}\label{aXbX}
|a_X\rangle &=& \cos\frac{\psi_a}{2}|X \uparrow + \rangle - e^{i\phi_a}\sin\frac{\psi_a}{2}|X \downarrow - \rangle, \\
|b_X\rangle &=& -\cos\frac{\psi_b}{2}|X \uparrow - \rangle + e^{i\phi_b}\sin\frac{\psi_b}{2}|X \downarrow + \rangle,\nonumber
\end{eqnarray}
where the $X$-dependence of $\psi_\nu$, $\phi_\nu$ is implicit.
The many-body state is therefore a hybridized spin-valley configuration, which may be represented in terms of two local spin-$1/2$ pseudospin fields ${\bf S}_a(X)$, ${\bf S}_b(X)$ encoded by the Euler angles $\psi_\nu\in[0,\pi]$, $\phi_\nu\in[0,2\pi]$:
\begin{equation}\label{SabDef}
{\bf S}_\nu=\frac{1}{2}\left(\sin\psi_\nu\cos\phi_\nu,\sin\psi_\nu\sin\phi_\nu,\cos\psi_\nu\right),
\end{equation}
where $\nu=a,b$. Note that in Ref. \onlinecite{MSF2014}, our focus was on the derivation of the ground state and we had assumed trivial phase factors in Eq. (\ref{aXbX}): $\phi_a=\phi_b\equiv\phi=0$. However, there is actually a manifold of degenerate ground states with an arbitrary global phase $\phi\neq 0$. This implies the existence of a gapless mode associated with a slowly varying twist of the angle $\phi$, consistent with Ref. \onlinecite{tdhfa} as will be discussed in more detail below.

We now allow for fluctuations in the collective variables $\psi_\nu({\bf r})$, $\phi_\nu({\bf r})$ [where ${\bf r}=(x,y)$] which vary slowly in space with respect to the magnetic length $\ell$. Assuming further that $g_0\sim e^2/\ell$ and hence is much larger than the other interaction scales (for $\alpha=x,y,z$, $g_\alpha\sim e^2 a_0/\ell^2$ with $a_0$ the lattice spacing \cite{Kharitonov_bulk}), a semi-classical approximation yields an effective Hamiltonian of the form
\begin{equation}\label{EvsSaSb}
H[{\bf S}_a({\bf r}),{\bf S}_b({\bf r})]=\sum_{\bf r}\left\{\frac{\rho_0}{2}\sum_{\alpha=x,y,z}\sum_{\nu=a,b}|\nabla S^\alpha_\nu|^2 +H_{loc}({\bf r})\right\}
\end{equation}
where $\rho_0\propto g_0$ is the pseudospin-stiffness, and $H_{loc}({\bf r})$ is a local term. The latter can be derived by
evaluating the expectation value of the microscopic Hamiltonian Eq. (\ref{Hmicro})
in a state of the form Eq. (\ref{aXbX}), with the label $X$ replaced by {\bf r}. Defining a local projector
\begin{equation}\label{Pdef}
\mathcal{P}({\bf r})=|a_{\bf r}\rangle\langle a_{\bf r}|+|b_{\bf r}\rangle\langle b_{\bf r}|\; ,
\end{equation}
the local energy term can be expressed as
\begin{equation}\label{Eloc_P}
H_{loc}({\bf r})=\sum_{\alpha=x,y,z}g_\alpha\left\{(Tr[\mathcal{P}({\bf r})\sigma_0\tau_\alpha])^2-Tr[(\mathcal{P}({\bf r})\sigma_0\tau_\alpha)^2]\right\},
\end{equation}
where $Tr$ is the trace of a $4\times 4$ matrix in the basis set by the 4 states $|\uparrow \pm \rangle$, $|\downarrow \pm \rangle$. Employing Eqs. (\ref{aXbX}) and (\ref{Pdef}), we obtain
\begin{widetext}
\begin{eqnarray}\label{Eloc}
H_{loc}({\bf r})&=& -[E_z-U(x)]\cos\psi_a({\bf r})
-[E_z+U(x)]\cos\psi_b({\bf r}) \\
&-& (g_z+3g_{xy})\cos\psi_a({\bf r})\cos\psi_b({\bf r}) - (g_z-g_{xy})\sin\psi_a({\bf r})\sin\psi_b({\bf r})\cos[\phi_a({\bf r})-\phi_b({\bf r})]\; . \nonumber
\end{eqnarray}
\end{widetext}
Note that since the physical parameters obey $g_z>0$ and $g_{xy}<0$, the coefficient of the last term is always negative. Indeed, this term arises from the ferromagnetic coupling between the $a$ and $b$ pseudospins in the $XY$ plane, and tends to lock the relative planar angle $\phi_-=\phi_a-\phi_b$ to $\phi_-=0$. In contrast, $H_{loc}$ does not contain any explicit dependence on the symmetric combination $\phi_+=\phi_a+\phi_b$, signifying the gapless nature of its fluctuations.

Inserting Eq. (\ref{Eloc}) with $\phi_-=0$ into Eq. (\ref{EvsSaSb}), and minimizing $H[{\bf S}_a({\bf r}),{\bf S}_b({\bf r})]$ with respect to the remaining collective fields $\psi_a({\bf r})$ and $\psi_b({\bf r})$, yields the static domain wall structure $\psi_a^0({\bf r}),\psi_b^0({\bf r})$ described in Ref.  \onlinecite{MSF2014}: in the bulk, $\psi_a^0=\psi_b^0=\psi$ where in the CAF phase ($E_z<E_z^c=2|g_{xy}|$) $\psi$ is a nontrivial canting angle \cite{Kharitonov_bulk} obeying $\cos\psi=E_z/E_z^c$, and in the FM phase ($E_z>E_z^c$) $\psi=0$; the angles smoothly change towards the edge where $\psi_a^0=-\pi$, $\psi_b^0=0$ in both phases. Close to the CAF/FM transition ($E_z\to E_z^c$), the effective width of the domain wall is given by the diverging length scale
\begin{equation}\label{xi_def}
\xi\sim \sqrt{\rho_0/|E_z-E_z^c|}\; .
\end{equation}

To describe the dynamics of {\it quantum} fluctuations in the collective pseudospin fields compared to their ground state configuration, we next construct a path-integral formulation \cite{SpinBooks} in terms of the Euclidean action
\begin{equation}\label{S_2D}
\mathcal{S}_{2D}=\int_0^\beta d\tau \left\{-\frac{i}{2}\sum_{\bf r}\sum_{\nu=a,b}\cos\psi_\nu\partial_\tau\phi_\nu+H[{\bf S}_a,{\bf S}_b]\right\},
\end{equation}
where $\beta=1/T$, $H[{\bf S}_a,{\bf S}_b]$ is given by Eq.
(\ref{EvsSaSb}), and the local fields are now $\psi_\nu({\bf
r},\tau)$, $\phi_\nu({\bf r},\tau)$ with $\tau$ the imaginary
time; here we have used units where $\hbar=k_B=1$. Defining the
fluctuation fields $\Pi_\nu({\bf r},\tau)$ via the substitution
\begin{equation}\label{Pi_def}
\cos\psi_\nu=\cos\psi_\nu^0+\Pi_\nu
\end{equation}
in the first term of Eq. (\ref{S_2D}), it is apparent that $\Pi_\nu$ are the canonical momenta of the planar angle fields $\phi_\nu$. Employing the canonical transformation into symmetric and antisymmetric fields
\begin{eqnarray}\label{ab2+-}
\phi_+ &=&\frac{1}{2}\left(\phi_a+ \phi_b\right)\,,\quad \phi_-=\phi_a- \phi_b, \\
\Pi_+ &=&\Pi_a +\Pi_b\,,\quad \Pi_-=\frac{1}{2}\left(\Pi_a- \Pi_b\right), \nonumber
\end{eqnarray}
the effective action acquires the form
\begin{eqnarray}\label{S_2D_pm}
&&\mathcal{S}_{2D}= \\
&&\int_0^\beta d\tau \left\{-\frac{i}{2}\sum_{\bf r}\sum_{\mu=+,-}\Pi_\mu\partial_\tau\phi_\mu+H[\Pi_+,\Pi_-,\phi_+,\phi_-]\right\},\nonumber
\end{eqnarray}
where in the last term, the dependence on $\phi_+$ is restricted to gradient terms, while the $\phi_-$-dependence includes a mass term [the last term in Eq. (\ref{Eloc}), $\propto\cos\phi_-$] independent of $E_z$. As a result, the normal modes of the antisymmetric sector are typically gapped, and a low-energy effective field-theory model can obtained by projecting to the symmetric sector encoded by the pair of conjugate fields $\phi_+,\Pi_+$. We note that
the local momentum operator $\Pi_+$, denoting a fluctuation in the total spin component $S_z$,
\begin{equation}\label{Pi+Sz}
\Pi_+=\delta S_a^z+\delta S_b^z=\delta S^z\; ,
\end{equation}
commutes with all the local terms of
$H[\Pi_+,\Pi_-,\phi_+,\phi_-]$. As we show in the next sections,
in the FM phase this leads to the emergence of a gapless edge mode which carries
fluctuations in $\phi_+$ (physically representing rotations of the
total spin in the $XY$ plane), and is protected by an approximate
conservation of the spin component $S^z$ in the edge sector.

\section{Normal modes and Effective Model}
\label{sec:normalmodes}
The low-energy dynamics of the model
discussed in the previous section is complicated by the fact that
the ground-state of the system is non-uniform in the $\hat{x}$
direction due to the edge potential.  In the FM phase, there are
gapless low-energy excitations which are confined to the edge of
the system \cite{Fertig2006}, whereas all excitations in the bulk
are gapped. As described above we are primarily interested in transport
due to the low-energy edge excitations and how this is impacted by
the bulk excitations at low but finite temperature. Accomplishing
this involves the challenge of developing a theory which includes
both the edge and bulk excitations, and interactions between them.
A natural description of the edge modes involves tilting the spin
orientations away from their semiclassical groundstate, for
example using the degrees of freedom $\phi_{\pm}$ and their
conjugates $\Pi_{\pm}$ in Eq. (\ref{ab2+-}). As argued in the last
section, only gradient terms of the variable $\phi_+$ can appear
in the effective action, Eq. (\ref{S_2D_pm}), leading to gapless
modes which will dominate the low-temperature transport properties
of the system \cite{Fertig2006,Kharitonov_edge}.

The difficulty with using this parameterization for the entire
system lies in the rather different orientations of the spins in
the semiclassical ground-state configuration near the edge and
deep in the bulk.  The problem is apparent in Eq. (\ref{Pi_def}).
In the FM state, deep in the bulk spins are oriented along the
$\hat{z}$ direction; i.e., $\psi_{\nu}^0=0$.  This means that one
should restrict $\Pi_{\nu}<0$ for fluctuations that are physically
allowable: spins can only fluctuate {\it downward} from this
orientation. Such a constraint is very challenging to implement in
a fluctuating field theory. One may prefer in this situation to
retain the original spin variables, $\vec S_\nu$, for which
$\langle S_\nu^z \rangle=1/2$ in the ground-state, and
$S_\nu^{x}$, $S_\nu^{y}$ are conjugate variables. This is just the
standard approach to spin waves \cite{SpinBooks}.

Thus, there is an essential tension between the natural degrees of
freedom in the bulk and at the edge.  In this section we will
introduce an effective model in which we write both the bulk and
the edge degrees of freedom in their ``natural'' representations,
while retaining the basic symmetries of the system, and thereby
introducing couplings that will allow energy to be exchanged
between the bulk and the edge.

\subsection{Single Component Model: Ground state}
We begin first with a simplified model meant to represent only the
lowest energy degrees of freedom of the system, which captures
both the variation of the spins at the edge and the change in the
gapless mode structure as the system passes through the CAF-FM
transition, but is simple enough to allow analytic progress to be
made.  By developing this model we will be able to gain insight
into how parameters of our effective model should behave. Towards
this end we introduce the energy functional
\begin{equation}
{E}[\hat n] = \int_{x>0} d^2r\left\{-E_zn_z + \tilde g n_z^2 +
\frac{\rho_0}{2}\sum_{\alpha=x,y,z}|\vec\nabla n_{\alpha}|^2
\right\}, \label{energy_func}
\end{equation}
where $n({\bf r})$ is a unit vector field
($\sum_{\alpha}n_{\alpha}({\bf r})^2=1$) on the two-dimensional
domain ${\bf r}=(x>0,y)$. Qualitatively, one could identify this
degree of freedom with the spin-$1$ field obtained from the
symmetric combination ${\bf S}={\bf S}_a+{\bf S}_b$ of the
spin-$1/2$ fields described in Section \ref{sec:2Daction}. Eq.
(\ref{energy_func}) is essentially a low-energy approximation of
the model given by Eqs. (\ref{EvsSaSb}) and (\ref{Eloc}) [with
$\tilde g \sim |g_{xy}|$], where $U(x)$ is replaced by a sharp boundary
condition at $x=0$, and ${\bf S}_a$, ${\bf S}_b$ are assumed to
obey the bulk condition $\psi_a({\bf r})=\psi_b({\bf r})$,
$\phi_a({\bf r})=\phi_b({\bf r})$ for all ${\bf r}$ except very close to the boundary.
This model supports two phases in its bulk, a
ferromagnet ($n_z=1$) for $E_z>E_z^c \equiv 2\tilde g$, and a
canted state ($n_z=E_z/2\tilde g$) for $E_z < E_z^c$.  To mimic
the behavior of the $\nu=0$ system edge, we impose the boundary
condition $n_z(x=0)=-1$, which forces a domain wall (DW) at the
edge into the groundstate configuration. In the FM state, the DW
configuration may be found analytically with standard techniques
\cite{rajaraman_book}.  Assuming a classical groundstate in which
the unit vector rotates through the $\hat x$ direction in going
from the bulk to edge, one writes $n_z(x) \equiv \cos\theta(x)$,
$n_x(x) \equiv \sin\theta(x)$, and the configuration $\theta(x)$
that minimizes the energy functional satisfies
\begin{equation}
\label{DW_eq_motion} \rho_0 \frac{d^2\theta}{dx^2} = E_z
\sin\theta - \tilde g \sin 2\theta.
\end{equation}
This is equivalent to the equation of motion for a particle at ``position'' $\theta$ accelerating
with respect to ``time'' $x$ through a potential
$$
V[\theta] = E_z \cos\theta - {1 \over 2} \tilde g \cos 2\theta.
$$
Assuming the system is in the FM state in the bulk, we must have
$\theta \rightarrow 0$ as $x \rightarrow \infty$, which fixes the
total energy of the fictitious particle at $E_z - \tilde g/2$.
Using energy conservation one then finds that the particle
``velocity'' obeys the equation
\begin{equation}
\label{DW_velocity}
\frac{d\theta}{dx} =
-\frac{\left[E_z(1-\cos\theta) - {1 \over 2} \tilde g (1-\cos 2\theta) \right]^{1/2}}{\sqrt{\rho_0/2}}.
\end{equation}
Equation (\ref{DW_velocity}) may be recast in an integral form
$$
\frac{x}{\sqrt{\rho_0/2}} =
-2 \int_{\pi/2}^{\theta(x)/2}
\frac{d\psi}{\sin\psi \left[ 2E_z -4\tilde g \cos^2 \psi \right]^{1/2}},
$$
for which the integral may be computed explicitly.  Defining the length scale
\begin{equation}
\ell_{DW} \equiv \sqrt{\frac{\rho_0/2}{2E_z-4\tilde g}}=\sqrt{\frac{\rho_0}{4(E_z-E_z^c)}},
\end{equation}
which is clearly the analog of $\xi$ [Eq. (\ref{xi_def})], this
leads to the equation
\begin{widetext}
\begin{equation}
z \equiv e^{-x/\ell_{DW}} =
\left[
\frac{1-\cos\theta/2}{1+\cos\theta/2}
\right]
\frac{E_z + 2\tilde g\cos\theta/2+ \sqrt{E_z-2\tilde g} \sqrt{E_z - 2\tilde g \cos^2\theta/2}}
{E_z - 2\tilde g\cos\theta/2+ \sqrt{E_z-2\tilde g} \sqrt{E_z - 2\tilde g \cos^2\theta/2}}\,.
\label{zsq}
\end{equation}
Finally, Eq. (\ref{zsq}) may be inverted, which (using the boundary conditions on $\theta$)
yields the result $\cos\theta(x) = 2y^2(x) +1$, with
\begin{eqnarray}
\label{ysq}
y^2(x)=
\frac{1}{2\left[2\tilde g (1-z)^2+r^2(1+z)^2\right]}
\Biggl\lbrace
r^2(z+1)^2+(E_z+2\tilde g)(z-1)^2
\quad\quad\quad\quad\quad\quad\quad\quad  \\
-\left[\bigl(r^2(z+1)^2+(E_z+2\tilde g)(z-1)^2\bigr)^2
-4E_z(z-1)^2\bigl(2\tilde g (1-z)^2 + r^2(1+z)^2 \bigr) \right]^{1/2}
\Biggr\rbrace, \nonumber
\end{eqnarray}
where the quantity $r \equiv \sqrt{E_z - E_z^c}$ measures how close the system
is to the transition between the FM and canted phases.
\end{widetext}

The DW in this model is essentially analogous to what was found in the edge DW for
the $\nu=0$
FM state discussed in Section \ref{sec:2Daction}.   At distances from the edge
larger that $\ell_{DW}$, $\theta(x)$ becomes
very small, and approaches zero (the bulk value for the FM state) exponentially,
$\theta(x) \sim e^{-x/2\ell_{DW}}$.  One can also solve for the DW shape exactly at the
critical value $E_z=E_z^c$, either using the method above or by taking the $r \rightarrow 0$ limit
of Eq. (\ref{ysq}).  The result is
\begin{equation}
\label{r_eq_0_DW}
\theta(x) \rightarrow \theta_c(x) \equiv 2 {\rm arccot}\left[\sqrt{\frac{\tilde g}{\rho_0}}x\right].
\end{equation}

\subsection{Single Component Model: Fluctuations}
We next consider the normal modes around this classical energy minimum.  A simple way to
proceed is to define a unit vector $\hat n^{\prime}(x)$ such that $n^{\prime}_z(x)=1$ in the
classical groundstate. This is accomplished by taking $n_y^{\prime}=n_y$, and
\begin{eqnarray}
\left(
\begin{array}{c}
n_x(x) \\
n_z(x)
\end{array}
\right)
=
\left(
\begin{array}{c c}
\cos\theta_{DW}(x) & \sin\theta_{DW}(x) \\
-\sin\theta_{DW}(x) & \cos\theta_{DW}(x)
\end{array}
\right)
\left(
\begin{array}{c}
n_x^{\prime}(x) \\
n_z^{\prime}(x)
\end{array}
\right),
\nonumber
\end{eqnarray}
where $\theta_{DW}(x)$ is the DW configuration which minimizes the energy functional.
Substituting this into Eq. (\ref{energy_func}), and writing
$n_z^{\prime} = \sqrt{1-n_x^{\prime 2}-n_y^{\prime 2}} \approx
1-(n_x^{\prime 2}+n_y^{\prime 2})/2$, after some algebra one arrives
at an energy functional which may be written to quadratic order
in the form
\begin{equation}
{H}[\hat n^{\prime}] \approx \sum_{\mu=x,y}\int_{x>0} d^2r
\left\{ n^{\prime}_{\mu}({\bf r}) \left[ -\frac{\rho_0}{2} \nabla^2 + U_{\mu}(x)
\right] n^{\prime}_{\mu}({\bf r}) \right\} ,
\label{E_quad}
\end{equation}
with ``potentials''
\begin{eqnarray}
\label{eq:UxUy}
U_x(x) &=& {1 \over 2}E_z \cos\theta_{DW}(x)-\tilde g \cos 2\theta_{DW}(x),  \\
U_y(x) &=& {3 \over 2}E_z \cos\theta_{DW}(x) -2\tilde g \cos^2\theta_{DW}(x) - E_z + \tilde g. \nonumber
\end{eqnarray}

To obtain the normal modes from this, it is convenient to impose angular momentum
commutation relations on the components of the unit vector,
$[n_x^{\prime}({\bf r}_1),n_y^{\prime}({\bf r}_2)] = 2i\delta({\bf r}_1 - {\bf r}_2) n^{\prime}_z({\bf r}_1)
\approx 2i\delta({\bf r}_1 - {\bf r}_2)$.  The last step, in which $n_z^{\prime}$ is replaced
by its groundstate value of 1, is the spin-wave approximation \cite{SpinBooks}.

The classical groundstate we have chosen in assuming the DW rotates through the
$n_x - n_z$ plane is a broken symmetry state of Eq. (\ref{energy_func}); globally rotating the
unit vector configuration around the $n_z$ axis yields a different configuration with
exactly the same energy.  Because of this, the quadratic Hamiltonian [Eq. (\ref{E_quad})] must
host a zero mode \cite{rajaraman_book}.  This can be directly identified with an eigenfunction
of the operator $-\frac{\rho_0}{2} \nabla^2 + U_y(x)$ with zero eigenvalue, $S_0(x)$,
where $S_0(x) \equiv \sin\theta_{DW}(x)$.  Note that this zero mode is confined to
the region of the domain wall, and is independent of the real-space coordinate $y$.

Because the Hamiltonian
and groundstate are uniform in the $\hat y$ direction, the normal modes have
well-defined momentum $q_y$.  One may exploit this by writing $n_x^{\prime}({\bf r})=
\int dq_y m_x(x,q_y)e^{iq_yy}/\sqrt{2\pi}$ and $n_y^{\prime}({\bf r})= \int dq_y
m(x,q_y)e^{-iq_yy}/\sqrt{2\pi}$. The normal mode Hamiltonian may now be written
as
\begin{eqnarray}
H &=&\int d{q_y}\int_{0}^\infty dx \Bigl\{m_x(x,-q_y)\left[h_x(q_y)+{1 \over 2}\rho_0q_y^2\right]m_x(x,q_y) \nonumber \\
&+& m_y(x,-q_y)\left[h_y(q_y)+{1 \over 2}\rho_0q_y^2\right]m_y(x,q_y)\Bigr\},
\label{sw_ham}
\end{eqnarray}
with operators $h_{\mu}=-{1 \over 2}\rho_0\partial_x^2 + U_{\mu}(x)$.
Note that we expect
the effectively one-dimensional operators $m$ to obey
$m_{x,y}(x,q_y)=m_{x,y}(x,-q_y)^{\dag}$.
The normal modes of
the system are determined by the eigenvalues and eigenfunctions of the operators $h_{x,y}$,
which are difficult to determine analytically.  The potentials associated with them,
$U_{x,y}(x)$ [Eq. (\ref{eq:UxUy})], both reach the constant value of $E_z/2 -\tilde g$ at large positive $x$.
Thus these operators will have continuous spectra of eigenvalues above this energy scale, which
becomes the frequency edge for spin-waves in the bulk of the system.

\subsection{Bulk Hamiltonian}

If we wish to focus on the behavior deep in the bulk, one can simply set
$U_{x,y}(x) \rightarrow E_z/2 -\tilde g$ and extend the domain of $x$ to
$-\infty < x < \infty$.  The resulting bulk Hamiltonian can be written
\begin{eqnarray}
H_{b}=\frac{1}{2}\sum_{\alpha=x,y}\int d^2r \Bigl[ S_{\alpha}({\bf r})
\left(E_z -2\tilde g -\rho_0 \nabla^2 \right)S_{\alpha}({\bf r}) \bigr], \nonumber \\
\label{bulk_ham_sw}
\end{eqnarray}
where we have made the identification
$ m_\alpha({\bf r}) \equiv  S_\alpha({\bf r})$,
the components of the unit vector in real space.
$H_{b}$ supports a gapped spin-wave mode of frequency $\omega(q) = E_z -2\tilde g +\rho_0 q^2$
which becomes gapless at the phase transition, i.e. when $E_z$ acquires the critical value $E_z^c=2\tilde g$; this behavior is highly
analogous to what is found for the low energy modes in the full system near the
transition in time-dependent Hartree-Fock calculations \cite{tdhfa}. Alternatively,
one may rewrite Eq. (\ref{bulk_ham_sw}) in terms of bosonic raising and lowering operators,
$a({\bf r})=(S_x({\bf r})+iS_y({\bf r}))/\sqrt{2}$,
$a^{\dag}({\bf r})=(S_x({\bf r})-iS_y({\bf r}))/\sqrt{2}$, which upon taking the
ferromagnetic groundstate average for the $S_z$ component of the spin \cite{SpinBooks}
yields the
needed commutation relations $[a({\bf r}),a^{\dag}({\bf r}')]=\delta({\bf r}-{\bf r}')$,
so that
\begin{widetext}
\begin{equation}
H_b= \int d^2r \left[ -{1 \over 2}\rho \left(
a^{\dag}({\bf r}) \nabla^2 a({\bf r}) +
a({\bf r}) \nabla^2 a^{\dag}({\bf r}) \right)
+\Delta a^{\dag}({\bf r}) a({\bf r}) \right].
\label{bulk_ham_bos}
\end{equation}
\end{widetext}
In writing Eq. (\ref{bulk_ham_bos}) we have dropped the subscript $0$ in $\rho_0$,
and
\begin{equation}
\Delta=E_z-2\tilde g=E_z-E_z^c\; .
\label{DeltaDef}
\end{equation}
Note that we expect $H_b$ more generally to be the long-wavelength form of the Hamiltonian
governing the low energy modes deep in the bulk of the FM state of the $\nu=0$ quantum Hall
state, with $\Delta \rightarrow 0$ as the transition to the CAF state is approached.

\subsection{Edge Hamiltonian}
\label{sec:Edge_Hamiltonian}
We next turn to a discussion of the lowest energy mode of the FM phase, which
as discussed above is a gapless edge state mode.
Near the edge,
$U_y$ of Eq. (\ref{eq:UxUy}) has a well potential which monotonically increases with increasing $x$ towards its
asymptotic value.  It is also interesting to note that $U_x = U_y + \Delta U$, with
$\Delta U = E_z[1-\cos\theta_{DW}(x)] \ge 0$, so that $U_x(x) \ge U_y(x)$ for any $x$.
Presuming our domain wall structure is stable, there cannot be any negative energy
states associated with either $h_x$ or $h_y$ in Eq. (\ref{sw_ham}).  We have seen that for $q_y=0$, $h_y$
supports a zero energy state; this is unlikely to be the case for $h_x$ because the effective
potential associated with it is larger than that of $h_y$.  It is possible that there are
bound states in the spectral interval $[0,E_z/2 -\tilde g]$, but as the critical
value of $E_z$ is approached this becomes a very small interval and so is unlikely to
host any bound states.  Thus we assume that there is only one bound state in the spectra
of $h_x$ and $h_y$ for $q_y=0$, associated with $h_y$, at zero eigenvalue.  With increasing
$q_y$ there will be a single linearly dispersing mode, which we associate with the gapless edge
excitation of the system.
Then the lowest energy modes of the system in the FM phase are the single gapless edge mode
and the bulk spin wave modes discussed in subsection C.
We note that the absence of other low-energy modes
is in apparent agreement with time-dependent Hartree-Fock results for the full $\nu=0$
spectrum \cite{tdhfa}.

In order to write down an effective Hamiltonian for the edge mode it is useful
to consider the equations of motion for $m_y(x)$ and $m_y(x)$.  Using
$\partial_t {\cal O} = i[H, {\cal O}]$, we find
\begin{eqnarray}
\partial_t m_x &=&4 (h_y+{1 \over 2}\rho_0q_y^2)m_y \nonumber \\
\partial_t m_y &=&-4 (h_x+{1 \over 2}\rho_0q_y^2)m_x. \nonumber
\end{eqnarray}
The two equations can be combined to give, after Fourier transforming with
respect to time,
\begin{equation}
\omega^2 m_y = 16(h_x+{1 \over 2}\rho_0q_y^2)(h_y+{1 \over 2}\rho_0q_y^2)m_y.
\label{2nd_order}
\end{equation}
For $q_y=0$ this equation is solved by $m_y=S_0(x)$ and $\omega=0$, and
we are interested in the solution that smoothly joins to this in the limit
$q_y \rightarrow 0$.  To quadratic order in $q_y$, this may
be written in real time as
$m_y(q_y,t) = [S_0(x)+\delta S(x,q_y)]\phi(q_y,t)$, with $\delta S$ of order $q_y^2$. Using
the fact that $h_yS_0(x)=0$, the equation of motion to order $q_y^2$ becomes
\begin{equation}
-\partial_t^2 S_0(x) \phi(q_y,t) =
16[h_xh_y \delta S(x,q_y) +{1 \over 2}\rho_0q_y^2 h_xS_0(x)]\phi(q_y,t).
\label{phi_eq_1}
\end{equation}
Recalling our assumption that $h_x$ does not support a zero mode,
it will have a well-defined inverse operator $h_x^{-1}$ which
we can apply to Eq. (\ref{phi_eq_1}).  Finally, multiplying the
whole equation by $S_0(x)$ on the left, integrating with respect
to $x$, and using the fact that
$\langle S_0 | h_y | \delta S \rangle \equiv \int_0^{\infty} dx S_0(x)h_y \delta S(x) = 0$
for any $\delta S$, we obtain the equation of motion
\begin{equation}
\left[-8\rho_0 q_y^2-\langle S_0| h_x^{-1} | S_0 \rangle\partial_t^2 \right]\phi(q_y,t) = 0.
\label{phi_eq}
\end{equation}
Thus we find a linearly dispersing normal mode $\omega(q_y) = u_0 q_y$,
with velocity $u_0=\sqrt{8\rho_0/\langle S_0| h_x^{-1} | S_0 \rangle}$.

The gapless edge mode obtained above is the only mode
in the FM phase that
approaches zero energy.  The variable $\phi$ represents an amplitude to
rotate the spins of the DW into the $\hat{y}$ axis from the
$\hat{x}$ axis through which we assumed the spins spatially rotate in the
classical DW groundstate.  Qualitatively, one may associate it with
an azimuthal angle of the spins at the center of the DW, and it plays
a role highly analogous to the $\phi_+$ degree of freedom in
Section \ref{sec:2Daction}.  Quantizing this degree of freedom
leads to a standard Luttinger liquid Hamiltonian, which, after
Fourier transforming into real space, may be
written in the form
\begin{equation}
H_e={u_{_{NM}} \over {2\pi}} \int dy \left\{ K_{_{NM}} \left(\pi \Pi(y) \right)^2
+{1 \over K_{_{NM}}} \left( \partial_y\phi(y) \right)^2 \right\},
\label{eq:LL_NM}
\end{equation}
with $[\Pi(y),\phi(y')]=-i \delta(y-y')$, and $u_{_{NM}}=u_0$.
Note that because $\Pi$ and $\phi$ are conjugate, the former can be
identified with deviations of spins near the center of the DW into the
$S_z$ direction, as expected from the general considerations of
Section \ref{sec:2Daction}.  Because the energy cost for spatial gradients
in $\phi$ descends directly from the two-dimensional spin stiffness $\rho_0$,
we expect $u_{_{NM}}/\pi K_{_{NM}} \sim \rho_0$, which remains finite and non-vanishing
even as the transition point is approached (i.e., when the gap Eq. (\ref{DeltaDef}) obeys $\Delta \rightarrow 0$).  This implies that the Luttinger
parameter behaves as
\begin{equation}
K^{-1}_{_{NM}} \sim \frac{\pi}{2} \left( \rho_0 \langle S_0 | h_x^{-1} | S_0 \rangle \right)^{1/2}.
\label{K_tot}
\end{equation}

Two comments are in order.  First, as explained in
Appendix \ref{sec:uK_critical}, $\langle S_0 | h_x^{-1} | S_0 \rangle
\sim \Delta^{-1/2}$, which is divergent as $\Delta \rightarrow 0$,
so that the Luttinger parameter $K$ vanishes in this limit.  This
means that the edge mode becomes extremely sensitive to perturbations
at the edge (in the renormalization group sense), so that the edge
Luttinger liquid cannot remain stable as the bulk transition from
a FM to a CAF phase is approached.  Secondly, the behavior of the
coefficients of the edge theory as the system approaches the
transition point are chosen to match what is found in the normal
mode theory.  Going beyond this to include coupling terms between
the bulk and edge modes will be most naturally accomplished by
writing them in a way that includes quadratic contributions, so
that this edge-bulk coupling leads to significant contributions to
the edge theory. Rather than deviate from our
development of our effective model, {we defer a fuller
discussion of this to Appendix \ref{sec:uK_critical}.}  At this point,
we introduce our model of the edge-bulk coupling.

\subsection{Bulk-Edge Coupling}

As described in Sec. \ref{sec:intro},
the gapless edge mode of this system
is in fact a helical, charge-carrying mode.  Spin waves described by the effective
one-dimensional theory above should be understood as carrying current in the
positive or negative direction, with amplitude proportional to the
deviation of the expectation value of $S_z$ in the excited state from
its groundstate value.  As in other topological systems \cite{TIreview}, dissipation
at zero temperature in this edge
system is then suppressed because backscattering requires spin-flip, which
cannot be accomplished by static disorder \cite{Fertig2006,SFP}.
At finite temperature, however, spin waves will always be present
in the bulk, so that
the edge system can exchange angular
momentum with it.

We thus introduce a phenomenological coupling which captures this process
and respects conservation of angular momentum, in the form
\begin{equation}
H_{int} = g\int dy \left\{ a^{\dag}(0,y)e^{i\phi(y)} + a(0,y)e^{-i\phi(y)}\right\}.
\label{H_int}
\end{equation}
Recalling that the bulk bosonic operators are actually spin raising and lowering
operators ($a=(S_x+iS_y)/\sqrt{2}$, $a^{\dag}=(S_x-iS_y)/\sqrt{2}$), one sees that the two terms
in $H_{int}$ respectively flip a spin down and up in the degrees of freedom
associated with $H_b$, at $x=0$, which is treated as the location of
the DW.  Compensating these spin flip operators are the operators
$e^{\pm i\phi(y)}$, which represent the opposing spin flips in the edge
system $H_{e}$.  This is easily understood when one recalls that the
$\Pi(y)$ operator represents the deviation of $S_z$ from its groundstate
configuration due to excitation of edge modes, and one may verify that
$e^{\pm i\phi(y)}$ are raising/lowering operators with respect to the
$\Pi$ operator \cite{Giamarchi}.  The two terms in $H_{int}$ thus each conserve $S_z$ in
the system as a whole ($H_b+H_e$).

Finally, we note that our full effective model, $H_b+H_e+H_{int}$, can
be expanded around a classical groundstate configuration to produce the normal
modes of the system. To be consistent,
modes deep in the bulk and at the edge should behave as $\Delta \rightarrow 0^+$
in the same way
as what we found for the
model introduced at the beginning of this section.  This analysis is discussed in more detail in App. \ref{sec:uK_critical}. It leads to the conclusion that the effective
``bare'' Luttinger parameter $K$ and spin wave velocity $u$
in $H_e$ scale with $\Delta$ in the same way as those in the normal mode theory, and the phenomenological constant $g$ vanishes with $\Delta$.  Specifically one finds
\begin{eqnarray}
u &\sim& \Delta^{1/4}, \nonumber \\
K &\sim& \Delta^{1/4}, \nonumber \\
g &\sim& \Delta^{3/4}. \nonumber \\
\end{eqnarray}
With this scaling one finds among
the normal modes for the fully coupled bulk-edge system
a gapless spin-wave mode, at the edge, with velocity scaling as $\Delta^{1/4}$,
as found for the simple model developed at the beginning of this section.

With this phenomenological model, we are now in a position to understand how
the coupling between the edge and bulk can impact transport in the ferromagnetic
state.

\section{Conductance}
\label{sec:G}

We now turn to the calculation of electric conductance, and investigate its dependence on temperature ($T$) and the Zeeman energy $E_z$. The results can be compared to the two-terminal conductance data of Ref. \onlinecite{Young2013}, and to potentially more systematic future studies at low $T$. We note that in both the CAF and FM phases, the lowest energy charged excitations are edge modes, and these are expected to dominate the d.c. electric transport. However, in the CAF the edge modes are still gapped, and the conductance at finite $T$ is therefore expected to exhibit an activated behavior of the form
\beq
G(E_z<E_z^c)\sim e^{-\Delta_c/T},
\label{eq:GCAF}
\eeq
where $\Delta_c$ has been shown \cite{MSF2014} to vanish when approaching the transition as $\Delta_c\sim (E_z^c-E_z)\log(E_z^c-E_z)$. We therefore
focus on the behavior in the FM phase, where the edge mode is gapless and naively one expects perfect conduction. Interestingly, as we show below, in this phase the {\it resistivity} at finite $T$ exhibits a similar activated form, reflecting a ``duality relation'' between the two phases.

Our starting point is the effective Hamiltonian derived in the previous section:
\begin{align}
\label{eq:Heff}
H_{eff} & =H_{e}+H_{b}+H_{int}\\
H_{e} & =\frac{u}{2\pi}\int\mathrm{d}y\left\{ K\left(\pi\Pi\right)^{2}+\frac{1}{K}\left(\partial_{y}\phi\right)^{2}\right\},\nonumber \\
H_{b} & =\int\mathrm{d}^2r\left\{- \frac{1}{2}\rho\left(a^{\dagger}\nabla^{2}a+a\nabla^{2}a^{\dagger}\right)+\Delta a^{\dagger}a\right\},\nonumber  \\
H_{int} & =g\int\mathrm{d}y\left\{ a^{\dagger}\left(0,y\right)e^{i\phi\left(y\right)}+a\left(0,y\right)e^{-i\phi\left(y\right)}\right\} , \nonumber
\end{align}
which describes a helical Luttinger liquid coupled
to a bath of 2D massive bosons along the line $x=0$. For simplicity, we assume here the 2D bulk to be an infinite plane rather than the semi-infinite plane $x>0$ considered in App. A: a straightforward calculation shows that the effect of bulk-edge coupling in the two cases is the same for an appropriate definition of the coupling constant $g$. The local bosonic fields $a({\bf r})$, $a^{\dagger}({\bf r})$ correspond, in the spin-wave approximation, to the bulk spin operators $S^-({\bf r})$, $S^+({\bf r})$, respectively; the canonically conjugate operators $\phi(y)$, $\Pi(y)$ encode, respectively, the planar angle and spin density $S^z_e(y)$ on the edge. We recall that the last term, representing the most relevant coupling between edge and bulk modes, can be traced back to a spin-flip term of the form $(S^+_bS^-_e + h.c.)$.

To the Hamiltonian describing the clean system Eq. (\ref{eq:Heff}), we next add a term which accounts for the coupling to a random potential associated with static impurities,
\beq
\label{eq:Hdis}
H_{dis}=-\int\mathrm{d}y\mu\left(y\right)\rho_e(y)=\frac{1}{\pi}\int\mathrm{d}y\mu\left(y\right)\partial_{y}\phi,
\eeq
where in the last step we have used the expression for the edge density operator in terms of the bosonic field $\phi$. Note that the helicity of the edge mode forbids standard backscattering terms [e.g. $\cos(2\phi)$] which would normally dominate the relaxation of charge current on the edge $j_e$ by direct coupling of left and right moving components. In the absence of coupling to the bulk via the term $H_{int}$ in Eq. (\ref{eq:Heff}), the edge mode thus obeys conservation of the total spin operator $\mathcal{S}^z_e=\int \mathrm{d}y\,S^z_e(y)$,
which is equivalent to the d.c. component of the charge current,
\beq
J_e=\int \mathrm{d}y\,j_e(y)=Ku\int \mathrm{d}y\,\Pi(y)\; .
\label{eq:J_edef}
\eeq
The forward scattering term Eq. (\ref{eq:Hdis}) can be absorbed into a redefinition of $\phi$ by the
transformation \cite{GS88,Giamarchi} $\phi\left(y\right) \rightarrow\phi\left(y\right)+(K/u)\int_{0}^{y}\mathrm{d}y'\mu\left(y'\right)$
leading to a random phase shift of the operators appearing in $H_{int}$:
\begin{align}
e^{i\phi(y)} & \rightarrow e^{i\phi(y)}\zeta\left(y\right)\; , \nonumber \\
\zeta\left(y\right) & \equiv
e^{i(K/u)\int_{0}^{y}\mathrm{d}y'\mu\left(y'\right)}\; .
\label{eq:zeta_def}
\end{align}
For a generic disorder potential, the random variable
$\zeta\left(y\right)$ can be assumed to satisfy \beq \left\langle
\zeta\left(y\right)\right\rangle _{dis}=0,\quad\left\langle
\zeta\left(y\right)\zeta^\ast\left(y'\right)\right\rangle
_{dis}=D\delta\left(y-y'\right), \label{eq:disorder} \eeq where
$\langle\dots\rangle_{dis}$ denotes an average over disorder.

The two-terminal conductance $G$ is next evaluated under the assumption that due to the almost conservation of $\mathcal{S}^z_e$ (and hence $J_e$) on each of the two edges, the intrinsic electric resistivity is small; i.e., in units of $e^2/h$,
\beq
G=\frac{2}{R_0+\delta R}
\label{eq:G2R}
\eeq
where $R_0\approx 1$ is the contact resistance arising from coupling of the leads to a single 1D channel, and $\delta R\ll 1$. Deviations of $R_0$ from the ideal value $R_0=1$ due to extrinsic processes (e.g., spin-relaxation in the contacts) reduces $G$ from the perfect $G=2$ value but may be assumed to have a negligible $T$-dependence. The intrinsic contribution $\delta R=L/\sigma$ (where $L$ is the length of the sample in the edge direction and $\sigma$ is the d.c. conductivity) is treated perturbatively in the rate of scattering.

To this end, we employ a hydrodynamic approximation \cite{forster} of the Kubo formula  for $\sigma$,
\beq
\sigma=\lim_{\omega\rightarrow 0}\frac{1}{L\omega}\int_0^\infty\mathrm{d}t\,e^{i\omega t}\langle[J_e(t),J_e(0)]\rangle
\label{eq:Kubo}
\eeq
(the $ee$ component of the conductivity matrix $\hat{\sigma}$ in a basis of current operators $\{J_p\}$), whereby it can be recast in terms of the inverse of a memory matrix $\hat{M}$, encoding relaxation rates:
\beq
\hat{\sigma}=\hat{\chi} [\hat{M}]^{-1}\hat{\chi}\; .
\label{eq:sigma2M}
\eeq
Here $\hat{\chi}$ is the matrix of static susceptibilities
\beq
{\chi}_{pq}= \frac{1}{L}\int_0^\beta\mathrm{d}\tau\,\langle J_p(\tau)J_q(0)\rangle\equiv (J_p|J_q)
\label{eq:chi_def}
\eeq
(describing an ``overlap" of the operators $J_p$, $J_q$), and $\hat{M}$ is determined by correlation functions of the force operators
\beq\label{eq:force}
F_p=\dot{J}_p=i[H,J_p]\; ;
\eeq
generally, the explicit form of $\hat{M}$ is quite complicated \cite{forster}, however in the case where $(F_p|J_p)=0$ it greatly simplifies and
\beqr
M_{pq}&=&\lim_{\omega\rightarrow 0}\frac{C_{pq}(\omega)-C_{pq}(\omega=0)}{i\omega}\; ,\nonumber \\
C_{pq}(\omega)&=&\frac{-i}{L}\int_0^{\infty} \mathrm{d}te^{i\omega t}\langle [F_p(t),F_q(0)]\rangle\; .
\label{eq:memory}
\eeqr
In Eqs. (\ref{eq:Kubo}) through (\ref{eq:memory}), $\langle ...\rangle$ denotes thermal expectation value at temperature $T$. It is apparent from Eq. (\ref{eq:sigma2M}) that the matrix elements of $\hat{\sigma}$ are dominated by slow modes, for which $F_p$ and hence the the matrix element $M_{pp}$ is small. In particular, the presence of a conserved operator $J_c$ which commutes with the Hamiltonian (i.e. $F_c=0$) leads to the divergence of any physical conductivity $\sigma_{pp}$ (and hence vanishing of the resistivity) provided the cross susceptibility $\chi_{pc}\not=0$; in such a case, the current $J_p$ is protected by the conservation law and can not decay \cite{RA}. When the conservation law is only approximate, one obtains a finite relaxation rate dominated by the small memory matrix element $M_{cc}$.

In our case, the approximate conservation law protecting the
charge current on each edge is $\mathcal{S}^z_e$, which is
identical to $J_e$ up to a constant prefactor [Eq.
(\ref{eq:J_edef})] \cite{SROG}. This justifies a diagonal version of Eq.
(\ref{eq:sigma2M}) and one obtains
\beq
\delta R=\frac{L}{\sigma}=\frac{LM_{ee}}{\chi_{ee}^2},
\label{eq:deltaR2M}
\eeq
where, for a Luttinger liquid, $\chi_{ee}$ is easily computed
\cite{Giamarchi} to yield a constant $\chi_{ee}=2uK/\pi$.
Employing Eq. (\ref{eq:memory}) for $p=q=e$ (and a standard
identity for the retarded correlation function) we get
\beq \delta
R=-\frac{1}{2(uK/\pi)^2}\int_0^{\infty} \mathrm{d}t\,t\Im
m\{\langle F_e(t)F_e(0)\rangle\}
\label{eq:deltaR2int}
\eeq
where, substituting Eq. (\ref{eq:Heff}) for the effective Hamiltonian,
\beq\label{eq:Fe}
F_e=i[H_{eff},J_e]=i[H_{int},J_e]\; ;
\eeq
in the last step we have used $[H_e,J_e]=0$. The intrinsic
resistivity is therefore dominated by processes whereby the edge
spin is relaxed into the 2D bulk. We finally introduce the
disorder potential by performing the phase shift Eq.
(\ref{eq:zeta_def}), so that $H_{int}$ acquires the form
\beq
H_{int} =g\int\mathrm{d}y\left\{\zeta(y)
a^{\dagger}\left(0,y\right)e^{i\phi\left(y\right)}+\zeta^\ast(y)a\left(0,y\right)e^{-i\phi\left(y\right)}\right\}.
\label{Hint_dis}
\eeq
Evaluating $\delta R$ from Eq. (\ref{eq:deltaR2int}) to leading order in $H_{int}$ [Eq.
(\ref{Hint_dis})], we modify the definition of angular brackets $\langle
...\rangle$ to include the disorder averaging
$\langle ...\rangle_{dis}$.

\begin{figure}
\includegraphics[scale=0.5]{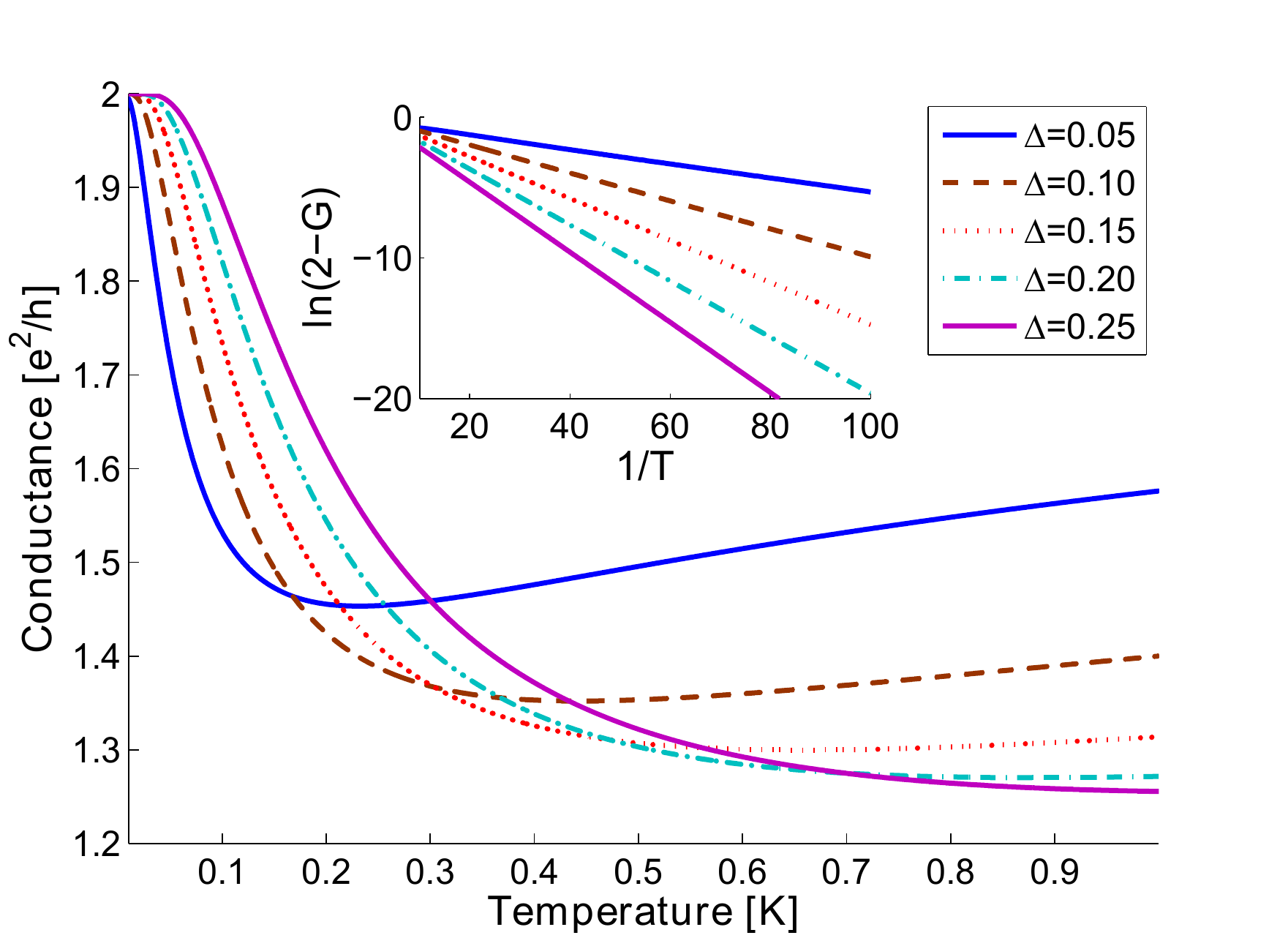}
\caption{(Color online.) Conductance in units of $e^2/h$ as a function of $T$, for different values of $\Delta$ in units of Kelvin. Assuming that $E_z^c\sim 1$K, we take $R_0=1$, $K=\Delta^{1/4}$, $u=u_0\Delta^{1/4}$ and $g=g_0\Delta^{3/4}$ where $u_0$, $g_0$ are such that the overall $\Delta$-independent prefactor of $\delta R$ [Eq. (\ref{eq:deltaR2B})] is 0.1. Inset: zoom on the low-$T$ regime $0.01$K$\leq T\leq 0.1$K. }
\label{fig:Conductance}
\end{figure}

We next employ Eqs. (\ref{eq:J_edef}), (\ref{eq:Fe}) and (\ref{Hint_dis}) to get the
correlation function
\begin{widetext}
\beqr
\langle F_e(t)F_e(0)\rangle &=&
(guK)^{2}\int\mathrm{d}y\int\mathrm{d}y'\left\langle
\zeta\left(y\right)\zeta^\ast\left(y'\right)\right\rangle
_{dis}\left\langle
e^{i\phi\left(y,t\right)}e^{-i\phi\left(y',0\right)} \left\{
a^{\dagger}\left(0,y,t\right)a\left(0,y',0\right)+
a\left(0,y,t\right)a^{\dagger}\left(0,y',0\right)\right\}\right\rangle
\nonumber \\ &\approx & (guK)^{2}D\int\mathrm{d}y\langle
e^{i\phi\left(y,t\right)}e^{-i\phi\left(y,0\right)}\rangle_e
\left\{ \langle
a^{\dagger}\left(0,y,t\right)a\left(0,y,0\right)\rangle_b+\langle
a\left(0,y,t\right)a^{\dagger}\left(0,y,0\right)\rangle_b\right\},\label{eq:FeFe}
\eeqr
\end{widetext}
where in the last step we have used Eq. (\ref{eq:disorder}), and
maintain the leading order in $g$ for which the thermal expectation
value is evaluated with respect to $H_0=H_e+H_b$ (where the bulk
and edge sectors are decoupled). Both sectors are described by free bosonic theories [see Eq. (\ref{eq:Heff})]. The edge part of the correlation function is given by the standard result for a Luttinger liquid \cite{Giamarchi},
\beq
\langle e^{i\phi\left(y,t\right)}e^{-i\phi\left(y,0\right)}\rangle_e =\lim_{\epsilon\rightarrow 0}\frac{\left(\frac{\pi\alpha}{\beta u}\right)^{K/2}\left(-1\right)^{-K/4}}{\left[\sinh\left(\frac{(t-i\epsilon)\pi}{\beta}\right)\right]^{K/2}},
\label{eq:LLcorr}
\eeq
where $\alpha$ is a short-distance cutoff. For the bulk, we use the bosonic correlation functions in momentum space
\beqr
\langle a^{\dagger}_{{\bf k}}(t)a_{{\bf k'}}(0)\rangle_b &=&e^{i\omega_{{\bf k}}t}n_{_B}({\bf k})\delta_{{\bf k},{\bf k'}}\; ,\nonumber \\
\omega_{{\bf k}} &=& \Delta+\rho |{\bf k}|^2,
\label{eq:acorr_k}
\eeqr
where $n_{_B}({\bf k})=1/(e^{\beta\omega_{{\bf k}}}-1)$ is the Bose function, and similarly
\beq
\langle a_{{\bf k}}(t)a^{\dagger}_{{\bf k'}}(0)\rangle_b =e^{-i\omega_{{\bf k}}t}[1+n_{_B}({\bf k})]\delta_{{\bf k},{\bf k'}}\; .
\eeq
The local correlation functions thus become
\beqr\label{eq:acorr_loc}
\langle a^{\dagger}\left(0,y,t\right)a\left(0,y,0\right)\rangle_b &=&\int\mathrm{d}^2k\,e^{i\omega_{{\bf k}}t}n_{_B}({\bf k}), \\
\langle a\left(0,y,t\right)a^{\dagger}\left(0,y,0\right)\rangle_b &=&\int\mathrm{d}^2k\,e^{-i\omega_{{\bf k}}t}[1+n_{_B}({\bf k})]\; .\nonumber
\eeqr
Inserting Eqs. (\ref{eq:acorr_loc}) and (\ref{eq:LLcorr}) into (\ref{eq:FeFe}) we obtain
for $\delta R$ [Eq. (\ref{eq:deltaR2int})]:
\begin{widetext}
\beqr
\label{eq:deltaR2C}
\delta R &\approx & -\mathcal{D}\int_0^{\infty} \mathrm{d}t\,t\Im m \{\mathcal{C}(t)\}\; , \quad {\rm where}\\
\mathcal{D} &\equiv & \frac{\pi^2g^2DL}{2}\left(\frac{\pi\alpha}{\beta u}\right)^{K/2}\; ,\quad
\mathcal{C}(t) \equiv \lim_{\epsilon\rightarrow 0}\frac{\left(-1\right)^{-K/4}}{\left[\sinh\left(\frac{(t-i\epsilon)\pi}{\beta}\right)\right]^{K/2}}
\int\frac{\mathrm{d}^2k}{(2\pi)^2}\left\{e^{i\omega_{{\bf k}}t}n_{_B}({\bf k})+e^{-i\omega_{{\bf k}}t}[1+n_{_B}({\bf k})]\right\}\; .\nonumber
\eeqr
\end{widetext}
Note that $D\propto n_{imp}$ where $n_{imp}$ is the density of
impurities per unit length; hence, the factor $DL$ encodes the
number of impurities $N_{imp}$. Performing the integrals in Eq.
(\ref{eq:deltaR2C}) we obtain (see Appendix \ref{sec:deltaRdetails}
for details)
\beqr
\delta R
&=&\frac{2^{\frac{K}{2}}\mathcal{D}\beta}{8\pi^2\rho}
\frac{\Gamma\left(1-\frac{K}{2}\right)\sin\left(\frac{\pi
K}{2}\right)}{2\pi}f(\beta\Delta)\; ,\nonumber \\
f(z) &\equiv &
\int_{z}^{\infty}\mathrm{d}x\left|\Gamma\left(\frac{K}{4}+i\frac{x}{2\pi}\right)\right|^{2}\frac{e^{-\frac{1}{2}x}}{1-e^{-x}}\;
. \label{eq:deltaR2B}
\eeqr
Recalling the $T$-dependence of
$\mathcal{D}$ [Eq. (\ref{eq:deltaR2C})], this yields $\delta R$ as
a function of $T$ for arbitrary values of the other parameters:
\beq
\delta R(T) \propto T^{\frac{K}{2}-1}f(\Delta/T)\; .
\label{eq:deltaR2T_short}
\eeq
In Fig. \ref{fig:Conductance}  we present $G$ vs. $T$ obtained directly from Eqs. (\ref{eq:G2R}) and (\ref{eq:deltaR2B}), for several values of $\Delta$ corresponding to a range of $E_z$ in the regime $E_z>E_z^c$.

We now consider the low $T$ limit where $T\ll\Delta$,
and use the asymptotic form of $f(z)$ at large argument to obtain the leading $T$-dependent contribution to the resistance (see App. \ref{sec:deltaRdetails}):
\beqr
\label{eq:deltaR_final}
\delta R(T) &\approx &R_{int}e^{-\Delta/T}\; , \\
R_{int}&\equiv &\frac{\pi\Gamma\left(1-\frac{K}{2}\right)\sin\left(\frac{\pi K}{2}\right)DL}{8\rho}\frac{g^2}{\Delta}\left(\frac{\alpha\Delta}{ u}\right)^{K/2},\nonumber
\eeqr
where we note that the prefactor of the exponential $R_{int}$ is $T$-independent. This simple activation of the resistance is remarkably reminiscent of the {\it conductance} in the CAF phase [Eq. (\ref{eq:GCAF})], where here the activation energy $\Delta\propto (E_z-E_z^c)$ [see Eq. (\ref{DeltaDef})] corresponds to the gap for spin-wave excitations in the bulk. Interestingly, the role it plays here is equivalent to a superconducting gap. The final expression for the low-$T$ two-terminal conductance in the FM phase is obtained by substituting Eq. (\ref{eq:deltaR_final}) into (\ref{eq:G2R}), yielding
\beq
G(E_z>E_z^c)\approx\frac{2}{R_0+R_{int}(\Delta)e^{-\Delta/T}},
\label{eq:GfinalFM}
\eeq
where the $E_z$-dependence is dominated by the behavior of $\Delta$ .

\section{Summary}
\label{sec:summary}

In this work we have developed an effective model for the ferromagnetic $\nu=0$
quantized Hall state of graphene, and used it to analyze the transport behavior
of the system at finite temperature.  The model includes a bulk system
supporting a gapped spin wave mode, an edge system supporting a charged
gapless helical mode, and a coupling term allowing an exchange of spin
between the two systems.
In principle the parameters of the effective theory which couples the edge and bulk are free. However, we use several ways to constrain them, especially in the ferromagnetic phase near the transition. We develop a simple nonlinear theory of the edge in the FM phase and match the $E_Z$-dependence of the spin-wave velocity of this model with the linear approximation to our effective theory.  We further make the physical demand that the spin stiffness should neither diverge nor vanish at the transition. This completely constrains the $E_Z$-dependence of all the free parameters of the effective theory.
An analysis in terms
of the memory matrix approach allows us to determine the temperature-dependence
of edge transport in this system.  In the presence of disorder, charged modes
of the system can be backscattered, with the necessary angular momentum for
such processes within a helical channel supplied by the bulk spin excitations.
This leads to concrete predictions for a two-terminal resistance
measurement of the system.

Our analysis leaves open a number of interesting further questions.  What is the
effect of disorder on the bulk of the system?  In particular, is there a range
of parameters for which gapless or nearly gapless spin excitations persist in
the bulk, leading to dissipative behavior over a broad range of temperatures
and/or Zeeman energies?  Our model can be easily generalized to capture the
canted antiferromagnetic phase, which is presumably seen as an insulating state
in experiments with relatively weaker Zeeman coupling.  Our approach in principle allows one
to compute the temperature dependence of transport in this phase as well.  More
challenging, and potentially very interesting, would be the transport
behavior of the system through the transition itself.  Connected to this,
it would be generally interesting to understand the bulk properties of
the system in the critical regime.  How the system behaves upon doping
is yet another interesting direction, an understanding of which would allow
further connection of our model with existing experimental data.
These and related questions will be addressed in future work.

\vskip 0.5in

{\it Acknowledgements -- } Useful discussions with E. Andrei, N. Andrei, T. Grover, P. Jarillo-Herrero, R. Shankar and A. Young
are gratefully acknowledged. The
authors thank the Aspen Center for Physics (NSF Grant No. 1066293) for
its hospitality. This
work was supported by the US-Israel Binational Science Foundation
(BSF) grant 2012120 (ES, GM, HAF), the Israel Science Foundation (ISF)
grant 231/14 (ES), by NSF Grant Nos. DMR 1306897 (GM), DMR-1506263 (HAF),
and DMR-1506460 (HAF).

\appendix

\section{Renormalization of edge parameters near criticality}
\label{sec:uK_critical}
In this Appendix we discuss some technical details that determine how
various parameters of our effective model scale with $\Delta$.  In particular
we demonstrate that the matrix element $\langle S_0 | h_x^{-1} | S_0 \rangle$
scales as $\Delta^{-1/2}$, as was stated in Section \ref{sec:Edge_Hamiltonian}.
We then discuss how this leads to the scaling of the parameters $u$, $K$, and
$g$ in our effective Hamiltonian.

\subsection{Small $\Delta$ behavior of $\langle S_0 | h_x^{-1} | S_0 \rangle$}
\label{sec:matrixelement}
We recall the operator $h_x \equiv -{1 \over 2}\rho_0\partial_x^2 + U_x(x)$, with
$$
U_x(x)={1 \over 2} E_z \cos\theta_{DW}(x) - \tilde g \cos 2\theta_{DW}(x),
$$
which has the asymptotic property $U_x(x\rightarrow \infty) = E_z/2-\tilde g
\equiv \Delta/2$.  As discussed in Section \ref{sec:Edge_Hamiltonian},
we assume for small $\Delta$ that $h_x$ has no bound states, in particular
no zero energy states, so that the operator $h_x^{-1}$ is well-defined.
The spectrum of $h_x$ then supports only scattering states, which
can be specified by eigenvalues of the form ${1 \over 2}\rho_0k_x^2+\Delta/2$,
with $k_x$ formally a continuous set of parameters labeling the spectrum.
Labeling the corresponding eigenvectors as $|k_x\rangle$, we then have
\begin{equation}
\langle S_0 |h_x^{-1}|S_0 \rangle = L_x\int_0^\infty {{dk_x} \over {2\pi}}
\frac{|\langle S_0 | k_x \rangle|^2}{{1 \over 2}\rho_0k_x^2+\Delta/2},
\label{matrix_element}
\end{equation}
where $L_x$ is a size scale which is taken to infinity in the thermodynamic
limit.
We next argue that the matrix element $\langle S_0 | k_x \rangle$
is finite for any $\Delta$, including at the critical value $\Delta=0$.
Since the wavefunctions $\psi_{k_x}(x)=\langle x | k_x \rangle$ are
increasingly unaffected by $U_x$ as $\Delta \rightarrow 0$, it is
sufficient to show that the matrix element is finite in this limit.
Because the wavefunctions in Eq. (\ref{matrix_element}) are normalized,
it is clear that the integrand is finite for large $k_x$ and that the
integral converges at its upper limit.  To see that there is no
divergence at the lower limit, we identify a length scale $\eta$ above
which the domain wall configuration $\theta_{DW}(x)$ is not appreciably
different than zero, so that for $x > \eta$ we can use an asymptotic
scattering form for $\psi_{k_x}(x)$, as well as
$S_0(x) \approx 2\sqrt{2\rho_0/\tilde g}/x \equiv \xi/x$ [see Eq. (\ref{r_eq_0_DW})].  Writing
$x=u/k_x$, the matrix element takes the form
$$
\langle S_0 | k_x \rangle \approx {{const.} \over {\sqrt L_x}}
+ {{\xi} \over {\sqrt{L_x}}} \int_{k_x\eta}^{\infty} du
\frac{e^{-iu} - e^{iu}e^{-2i\delta(k_x)}}{u},
$$
where $\delta(k_x)$ is the phase shift, which for small $k_x$
has the form $\delta(k_x) \approx -k_xa$, with $a$ the scattering
length.  It is clear from these forms that $\langle S_0 | k_x \rangle$
is finite in the limit $k_x \rightarrow 0$, and we write this limit
as $C/\sqrt{L_x}$.  Finally, noting that, for small $\Delta$,
Eq. (\ref{matrix_element}) is dominated by the lower limit on $k_x$,
we find
$$
\langle S_0 |h_x^{-1}|S_0 \rangle \sim \int_0^{\infty} dk_x
\frac{C^2}{\rho_0k_x^2 + \Delta/2} \sim 1/\sqrt{\Delta},
$$
which leads to the $u_{NM} \sim \Delta^{1/4}$ behavior discussed
in Section \ref{sec:Edge_Hamiltonian}.

\subsection{Scaling of $u$, $K$, and $g$ with $\Delta$}
\label{sec:uKg_scaling}
We next discuss how the parameters specifying the
one-dimensional part of our effective Hamiltonian, $u$ and $K$,
behave as the transition to the canted antiferromagnet (CAF) is
approached from the ferromagnetic (FM) side.  Our approach is specifically
to expand the Hamiltonian for small fluctuations around a classical
groundstate, and to specify the behavior of $u$ and $K$ to match
what was found in Section \ref{sec:normalmodes}.  We begin by rewriting the
effective Hamiltonian in the form
\begin{align}
\label{eq:Heff:appB}
H_{eff} & =H_{e}+H_{b}+H_{int},\\
H_{e} & =\frac{u}{2\pi}\int\mathrm{d}y\left\{ K\left(\pi\Pi\right)^{2}+\frac{1}{K}\left(\partial_{y}\phi\right)^{2}\right\},\nonumber \\
H_{b} & =\int\mathrm{d}^2r\left\{ \frac{1}{2}\rho\left(\vec\nabla a^{\dagger}\vec\nabla a+\vec\nabla a\vec\nabla a^{\dagger}\right)+\Delta a^{\dagger}a\right\},\nonumber  \\
H_{int} & =g\int\mathrm{d}y\left\{ a^{\dagger}\left(0,y\right)e^{i\phi\left(y\right)}+a\left(0,y\right)e^{-i\phi\left(y\right)}\right\} . \nonumber
\end{align}
The Hamiltonian has a global symmetry of the form $\phi(y) \rightarrow \phi(y) +
\varphi_0$, $a \rightarrow a e^{i\varphi_0}$, $a^{\dag} \rightarrow
a^{\dag} e^{-i\varphi_0}$.  This implies that classical groundstates form
a degenerate continuous manifold, and for convenience we consider fluctuations
around $\phi(y) = 0$.  For small but non-vanishing values of this field, to
quadratic order one finds
\begin{equation}
H_{int} \approx g \int dy \left\{[a+a^{\dag}][1-{1 \over 2}\phi(y)^2]
+i\phi(y)[a^{\dag}-a] \right\}.
\label{smallphi}
\end{equation}
Rewriting $a({\bf r}) \equiv  [P({\bf r}) + i Q({\bf r})]/\sqrt{2}$,
$a({\bf r})^{\dag} \equiv  [P({\bf r}) - i Q({\bf r})]/\sqrt{2}$ (i.e., $P$ and $Q$ denote the spin operators $S_x$ and $S_y$, respectively) with
$[P({\bf r}_1),Q({\bf r}_2)]=i\delta({\bf r}_1-{\bf r}_2)$, yields
\begin{equation}
H_{int} \approx \sqrt{2} g \int dy \left\{P(0,y)[1-{1 \over 2}\phi(y)^2]
+\phi(y)Q(0,y) \right\}.
\label{smallphiPQ}
\end{equation}
If $P$ is treated classically, it is clear that the Hamiltonian will be
minimized by $P({\bf r}) \ne 0$.  Collecting terms involving $P$ for $\phi=0$,
the function doing so will minimize
$$
H_P=\int_{x \ge 0^-} d^r \left\{ {1 \over 2} \rho | \vec\nabla P |^2
+ {1 \over 2} \Delta P^2 + \sqrt{2} g P(\bf r) \delta(x) \right\}.
$$
Minimizing this subject to the boundary condition $\partial_x P(x=0^-,y))=0$,
which is appropriate to an open boundary, one finds $P=P_0$ with
\begin{eqnarray}
P_0(x) &=& \frac{-\sqrt{2} g }{\sqrt{\rho\Delta}}  \quad\quad\quad  0^- < x \le 0, \nonumber\\
&=& \frac{-\sqrt{2} g }{\sqrt{\rho\Delta}}
e^{-\left(\frac{\Delta}{\rho}\right)^{1/2} x} \quad x > 0 \nonumber.
\end{eqnarray}
Writing $P=P_0+p$, the effective Hamiltonian at the quadratic level now has
the form
\begin{eqnarray}
H_{eff} &\approx& \int_{x>0^-} d^2r \left\{
{1 \over 2} \rho \left(|\vec\nabla p|^2 + |\vec\nabla Q|^2 \right)
+{1 \over 2} \Delta \left(p^2 + Q^2\right) \right\} \nonumber \\
&+& g \int dy \left\{
\frac{g}{\sqrt{\rho\Delta}} \phi(y)^2 + \sqrt{2} \phi(y) Q(0,y) \right\} \nonumber \\
&+& \frac{u}{2\pi} \int dy \left\{
K\left(\pi\Pi\right)^2 + {1 \over K} \left( \partial_y \phi \right)^2 \right\}. \label{AppA:quadratic}
\end{eqnarray}
The middle term in Eq. (\ref{AppA:quadratic}) encodes a coupling between the
$\phi$ and $Q$ fields, capturing the effects of the global symmetry described
above.  The effect of this coupling can be found explicitly by minimizing
Eq. (\ref{AppA:quadratic}) with respect to $Q$, subject to the boundary
condition $\partial_x Q(x=0^-,y)=0$, which again is appropriate for an
open boundary.  This minimum $\Phi(x,y)$ obeys the equation
$$
-\rho\nabla^2 \Phi + \Delta \Phi + \sqrt{2}g\phi(y)\delta(x) = 0\; .
$$
Fourier transforming with respect to $y$, the solution to this equation
for $x \ge 0$ is
$$
\Phi(x,q_y) = -\frac{\sqrt{2}g}{\rho}
\frac{\phi(q_y)}{\sqrt{\Delta + \rho q_y^2}}
e^{-\sqrt{(\Delta + \rho q_y^2)/\rho} \,x }.
$$
We can finally write $Q = \Phi+q$, with $[p({\bf r}_1),q({\bf r}_2)]=
i\delta({\bf r}_1-{\bf r}_2)$ to fully decouple the edge mode from the
bulk.  After some algebra, we arrive at the effective Hamiltonian at the quadratic level in the form
\begin{eqnarray}
\label{Heffquad:AppA} H_{eff} &\approx&
\int_{x \ge 0} d^2r \left\lbrace
{1 \over 2} \rho \left(|\vec\nabla p|^2 + |\vec\nabla q|^2 \right)
+{1 \over 2} \Delta \left( p^2 + q^2 \right) \right\rbrace
\nonumber \\
&+& \frac{u}{2\pi} \int dy \left\{
K\left(\pi\Pi\right)^2 + {1 \over K} \left( \partial_y \phi \right)^2 \right\}
\nonumber \\
&+& L_y \frac{g^2}{\sqrt{\rho\Delta}}\int \frac{dq_y}{2\pi}
\left[1 - \frac{1}{\sqrt{1+\rho q_y^2/\Delta}} \right]
\phi(-q_y)\phi(q_y). \nonumber\\
\end{eqnarray}

For small enough $q_y$, it is apparent that the last two terms of Eq.
(\ref{Heffquad:AppA}) support a linearly dispersing normal mode, whose dynamics is described by a Luttinger liquid Hamiltonian with renormalized parameters. In particular, the renormalized coefficient of the $(\partial_y \phi)^2$ term is $u/K+2\pi g^2 \sqrt{\rho}/\Delta^{3/2}$.  Our goal is to match
the Hamiltonian controlling this mode as $\Delta$ becomes small
to the result [Eq. (\ref{eq:LL_NM})] of the
model described in Section \ref{sec:normalmodes},
in which non-Gaussian properties
of the bulk system were retained.  This leads to two requirements:
({\it i}) The coefficient of the $(\partial_y \phi)^2$ should remain
finite and non-vanishing in the limit of small $\Delta$;
({\it ii}) the velocity of the gapless mode should vanish as $\Delta^{1/4}$.
The first condition will be met if we assume $g \sim \Delta^{3/4}$ and $u \sim K$. Noting further
that the product $uK$ [the coefficient of the $\left(\pi\Pi\right)^2$ in Eq.
(\ref{Heffquad:AppA})] is not renormalized,
requirement ({\it ii}) on the velocity implies that our ``bare'' parameters $u$ and $K$ scale
as $u \sim K \sim \Delta^{1/4}$, in accordance with the scaling of the normal
mode parameters $u_{NM}$, $K_{NM}$ derived in Section \ref{sec:normalmodes}.

\section{Derivation of the general expression for $\delta R$ vs. $T$}
\label{sec:deltaRdetails}

In this Appendix we first derive the general expression for $\delta R$ [Eq. (\ref{eq:deltaR2B})] starting from Eq. (\ref{eq:deltaR2C}). Inserting $\omega_{{\bf k}}$ from Eq. (\ref{eq:acorr_k}), writing the Bose function as a geometric sum and performing the integral over ${\bf k}$, the correlation function $\mathcal{C}(t)$ becomes
\begin{align}
\mathcal{C}\left(t\right)
&=\lim_{\epsilon\rightarrow0}\left(-1\right)^{-\frac{K}{4}}\left(\sinh\left(\frac{\left(t-i\epsilon\right)\pi}{\beta}\right)\right)^{-\frac{K}{2}} \label{eq:Ct_final} \\
&\times\frac{1}{4\pi\rho}\left(\sum_{n=0}^{\infty}\frac{e^{-\Delta\left(n\beta+it\right)}}{n\beta+it}+\sum_{n=1}^{\infty}\frac{e^{-\Delta\left(n\beta-it\right)}}{n\beta-it}\right). \nonumber
\end{align}
To proceed with the calculation of $\delta R$,
we recast Eq. (\ref{eq:deltaR2C}) as
\beqr
\delta R &\approx & -\frac{\mathcal{D}}{4\pi\rho}I \quad{\rm where} \nonumber \\
I &\equiv & 4\pi\rho \Im m \left\{\int_{0}^{\infty}\mathrm{d}t\,t\mathcal{C}(t)\right\}\; .
\eeqr
Substituting (\ref{eq:Ct_final}) for $\mathcal{C}\left(t\right)$, we get
\begin{widetext}
\begin{align}
I
&=\Im m\left\{ \int_{0}^{\infty}\mathrm{d}t\cdot t\left(-1\right)^{-\frac{K}{4}}\left(\sinh\left(\frac{t\pi}{\beta}\right)\right)^{-\frac{K}{2}}\left(\sum_{n=0}^{\infty}\frac{e^{-\Delta\left(n\beta+it\right)}}{n\beta+it}+\sum_{n=1}^{\infty}\frac{e^{-\Delta\left(n\beta-it\right)}}{n\beta-it}\right)\right\} \nonumber \\
&=\Im m\left\{ \int_{0}^{\infty}\mathrm{d}t\cdot t\left(-1\right)^{-\frac{K}{4}}\left(\sinh\left(\frac{t\pi}{\beta}\right)\right)^{-\frac{K}{2}}\int_{\Delta}^{\infty}\mathrm{d}\Delta'\left(\sum_{n=0}^{\infty}e^{-\Delta'\left(n\beta+it\right)}+\sum_{n=1}^{\infty}e^{-\Delta'\left(n\beta-it\right)}\right)\right\}
 \label{eq:I_details} \\
&=-\int_{\Delta}^{\infty}\mathrm{d}\Delta'\sum_{n=0}^{\infty}e^{-\Delta'n\beta}\frac{\partial F_{-}\left(\Delta'\right)}{\partial\Delta'}+\int_{\Delta}^{\infty}\mathrm{d}\Delta'\sum_{n=1}^{\infty}e^{-\Delta'n\beta}\frac{\partial F_{+}\left(\Delta'\right)}{\partial\Delta'} , \nonumber
\end{align}
where
\begin{align}
&F_{\mp}(\Delta)\equiv\Im m\left\{ \frac{\left(-1\right)^{-\frac{K}{4}}}{i}\int_{0}^{\infty}\mathrm{d}t\cdot e^{\mp i\Delta t}\left(\sinh\left(\frac{\pi}{\beta}t\right)\right)^{-\frac{K}{2}}\right\} =2^\frac{K}{2}\frac{\beta }{2\pi}\Im m\left\{ \frac{\left(-1\right)^{-\frac{K}{4}}}{i}B\begin{pmatrix}i\gamma\pm\frac{K}{4}, & 1-\frac{K}{2}\end{pmatrix}\right\} \nonumber \\
&=-2^\frac{K}{2}\frac{\beta }{2\pi}\Re e\left\{\Gamma\left(1-\frac{K}{2}\right)\left|\Gamma\left(\frac{K}{4}+i\gamma\right)\right|^{2}
\frac{\left(\cos\pi\frac{K}{4}-i\sin\pi\frac{K}{4}\right)}{\pi}\left(\cosh\pi\gamma\sin\frac{\pi K}{2}\mp i\sinh\pi\gamma\cos\frac{\pi K}{2}\right)\right\}  \label{eq:Fmp}\\
&=-2^\frac{K}{2}\frac{\beta }{2\pi}\Gamma\left(1-\frac{K}{2}\right)\left|\Gamma\left(\frac{K}{4}+i\gamma\right)\right|^{2}\frac{1}{\pi}e^{\mp\pi\gamma}\frac{1}{2}\sin\frac{\pi K}{2} ; \nonumber
\end{align}
\end{widetext}
here $\gamma =\frac{\beta \Delta}{2\pi}$,
$B(x,y)=\frac{\Gamma(x)\Gamma(y)}{\Gamma(x+y)}$ is the Beta function and we have used the identity $\Gamma(z)\Gamma(1-z)=\pi/\sin(\pi z)$. Inserting these expressions for $F_{+}$ and $F_{-}$ in Eq. (\ref{eq:I_details}) yields
\begin{align}
I
&=2^\frac{K}{2}\frac{\beta }{2\pi}\Gamma\left(1-\frac{K}{2}\right)\frac{1}{2\pi}\sin\frac{\pi K}{2}\times \label{eq:I_final} \\
&\Bigg\{-\int_{\Delta}^{\infty}\mathrm{d}\Delta'\frac{1}{1-e^{-\Delta'\beta}}\frac{\partial}{\partial\Delta'}
\left(\left|\Gamma\left(\frac{K}{4}+i\gamma'\right)\right|^{2}e^{-\frac{\beta\Delta'}{2}}\right) \nonumber\\
&+\int_{\Delta}^{\infty}\mathrm{d}\Delta'\frac{e^{-\Delta'\beta}}{1-e^{-\Delta'\beta}}\frac{\partial}{\partial\Delta'}
\left(\left|\Gamma\left(\frac{K}{4}+i\gamma'\right)\right|^{2}e^{\frac{\beta\Delta'}{2}}\right)\Bigg\}.\nonumber
\end{align}
Finally, after integration by parts we obtain the expression of Eq. (\ref{eq:deltaR2B}).

The asymptotic form of $f(z)$, which dominates the limit $T\ll \Delta$ (namely, $z \rightarrow \infty$), is now obtained from Eq. (\ref{eq:deltaR2B}) by substituting the asymptotic form of $\Gamma(z)$ at large arguments:
\beq
f(z)\approx\left(2\pi\right)^{2-\frac{K}{2}}\int_{z}^{\infty}\mathrm{d}x\, x^{\frac{K}{2}-1}e^{-x}\; .
\label{eq:fz_approx}
\eeq
It is therefore proportional to the incomplete Gamma function $\Gamma (\frac{K}{2},z)$, which can be further approximated for $z \rightarrow \infty$ to give
\begin{align}
f\left(z\right)\approx\left(2\pi\right)^{2-\frac{K}{2}}z^{\frac{K}{2}-1}e^{-z}\; .
\label{eq:fz_approx}
\end{align}
This leads to the approximate expression for $\delta R$ in Eq. (\ref{eq:deltaR_final}).

\end{document}